# Does Time exist? – catchphrases and concepts

*Fedde Benedictus*

*Utrecht University*

**Abstract**

This paper lends perspective to the media catchphrase that "time does not exist". To show how phrases such as this should be interpreted, we formulate some of the most important questions that arise in discussions about time in physics. We start by analyzing Newton's objective view on time and explain how entropy gives it a direction. After that, we discuss how relativity distances us from Newton's views, while quantum theory brings us closer to the Newtonian view. The arguments are presented in such a way that they can be followed and checked by anyone who studies the theories involved – be they undergraduate physics students or otherwise.

# Contents

## Two questions about time

Many popular-science books state that time does not exist[1], while physics textbooks are usually silent about the nature of time – about what they regard as philosophical questions. But one only has to look up to see that the statement that time does not exist is problematic. We see time all around us: from the rise of the sun to the change of the seasons, and from falling stones to decaying trees. Considering all this, how can one ever argue that time does not exist?

We are looking at the problem from the scientist's perspective, so we focus on the empirical (that which can be observed). What we observe in the case of falling stones and decaying trees is *change*, not time itself.

Rather than stating that time does not exist, we address the following two questions:

1. Is time more than change?[2]
2. Does time have a direction, and, if so, is the direction of time an objective matter?

The setup of this paper is as follows. First, we will discuss Newton's ideas about time and see how he answers the questions above. In this context we will encounter the ideas of *presentism*, *the direction of time*, and *probability*. After having acquainted ourselves with Newton's views, we will see how Einstein's relativity seems to wreak havoc with these ideas – lending credence to the idea of *eternalism*. In the subsequent chapter about quantum theory we will find, perhaps surprisingly, that the standard interpretation of quantum theory brings us closer to Newton's views.

## 1. Newton

Isaac Newton, considered as one of the fathers of modern science, had a clear view of time in physics. In his view, the universe in which we live is a 3-dimensional 'container' in which particles move around and clump together to form objects (like stars and planets). Time, in this picture, is a fixed background for the motion of particles within the 3D container – like a stage in a theatre.[3]

---

[1] Carlo Rovelli is often quoted as saying this [by The Guardian and The New Scientist, for example].
[2] At first sight, this might seem like a rhetorical question: since only change is observable, it would be against the spirit of the scientist to say that time is more than change. However, even the most staunch, empirically minded scientist would admit that there are some concepts and that are needed to make the scientific worldview coherent even though these concepts have no observable consequences (such as the concept of the moon when nobody is looking at her; we assume that the moon exists when nobody is looking at her to explain why we see the moon when we *do* look in her direction).
[3] Isaac Newton, *Principia* (1668).

## Time is more than change

In Newton's view time is clearly more than merely change since it is a fixed background. If there would be no change, there would still be a background – like an empty theatre stage that remains after the *exeunt omnes*.

However, it is not clear from this what time actually is. To illustrate this, consider the following: we have the sensation that the present is real, the future is not, and the past is... somewhere in between. How do we justify this? Why do we believe that the present is more real than the future? Are there *degrees of 'realness'*?

## Presentism/eternalism

A common (intuitively appealing) idea is that the possibility to causally interact with the present is what makes it real (we can *do* things in the present). The question how to define a 'present' has a straightforward answer in Newtonian physics. The possibility of (gravitational) influence to be transmitted over long distances instantaneously serves to define a global 'now': a point in time about which all observers can agree when it takes place. Newton's gravity as an 'action-at-a-distance' is what makes the idea that only the present is real intuitively appealing.

The view that only the present is real is called *presentism*, whereas the view that past, present and future are all equally real is called *eternalism*. There is also a midway view: the idea that the present and the past are real, but the future is not.[4]

## The Direction of Time

The three laws of Newton ([1] the law of inertia, [2] force equals mass times acceleration, [3] any action creates an equal and opposite reaction) are time-symmetric. A simple example will illustrate what we mean by that. If we see a YouTube video of two billiard balls bouncing off each other there is no way for us to tell whether we see the video as it was recorded or in reversed time. The same holds for any situation involving bouncing billiard balls: the same laws hold whether we look forwards or backwards in time. That is why we call these laws time-symmetric.

The situation is very different for many of the processes that we see around us. Take leaves falling from a tree or a glass falling to the ground and breaking in zillions of pieces. We would immediately spot it if the time of a video clip of such an event were reversed. And yet all the particles that constitute the leaves, the tree, the glass, and the table obey the time-symmetric laws of Newton.

## Paradox: whence time asymmetry?

We have a paradox here: how can it be that symmetric laws at the micro level (of the particles) give rise to an asymmetry in time at the macro level (of trees and leaves)? In other words, how can it be that time seems to have a direction?

---

[4] If interaction is all that counts, why not say that the future is real, but the past is not; our actions influence the future, but not the past. Or do they? Could it be empirically investigated whether the past can be causally influenced?

There is no consensus in the scientific community about how to answer this question, and we will not attempt to undo this Gordian knot. However, for the purpose of our discussion it is useful to formulate the question in terms of *entropy* because that allows us to see what is at stake here.

Before we give the definition of entropy, we must consider the concepts of the microstate and that of a macrostate. Those concepts are most easily understood when we consider a gas in a box. Any *macrostate* of that box is a description of the box in terms of macroscopic observables, such as pressure, temperature, and volume (*p*, *T* and *V*). A *microstate* of the box, on the other hand, is a description of the box in terms of all the molecules in the box (all their positions and velocities).

Given a certain macrostate, usually many microstates correspond to it. For example, consider a macrostate with a certain *p*, *T* and *V* and put the contents of the box upside-down in the box. *p*, *T* and *V* will be the same while the positions and velocities of the particles are different, showing how two different microstates can manifest themselves as the same macrostate. Many other rearrangements of the molecules also leave the values of *p*, *T* and *V* unchanged, and so give rise to the same macrostate. *The entropy of a specific macrostate is proportional to the number of microstates that can be associated with it.*[5]

### *Paradox solved: asymmetry due to entropy increase*
Does this solve our paradox? We asked how it can be that the time-symmetric laws that govern the gas-particles at the microscopic level give rise to an asymmetry in time at the macroscopic level. In terms of entropy our paradox becomes: why is it that entropy always increases?

Let us take another look at the gas in the box. If the gas-molecules can move around freely, it is much more likely that they spread evenly throughout the box than pile up in one of the corners of the box. This can be understood in terms of probability. If, at a certain moment, the density of particles in a certain part of the box is greater than in other parts of the box, the probability of gas-particles[6] colliding there (causing them to move to less densely populated parts of the box) is larger than particles colliding in less densely populated parts of the box. In time, therefore, the density will become approximately the same throughout the whole box; the gas will spread evenly throughout the box.

We can state this in terms of entropy: there are many more microstates associated with an even spread than there are associated with a situation in which all gas is piled up in one of the corners. This shows the working of *the second law of thermodynamics*: the entropy of a system always increases. It is important not to overstate our case here: we have not solved the paradox; we have not explained why time asymmetry exists. We have merely reformulated the fact that this symmetry exists in terms of entropy. However, our analysis *does* show us a way in which the paradox can be solved – by looking at the concept of probability.

---

[5] This definition is given by Boltzmann's equation: $S = k \cdot \log(W)$, which defines the entropy, $S$, of some physical system (some macrostate) as proportional to the logarithm of the number of possible microstates, $W$, of that system.
[6] This is a simplification as there are air-molecules with the gas-molecules in the box. The argument remains unchanged, however: We can ignore all collisions between gas-particles and air-molecules because on average they cancel each other out.

*Paradox renewed: what is entropy/probability?*

We have defined the direction of time in terms of entropy: the forward-direction in time is the time-direction in which entropy increases. But this is merely a restatement of our paradox. It shows *how* time symmetric microscopic laws give rise to an asymmetry in time at the macroscopic level, but it does not tell us *why* that happens. However, the definition[7] makes it possible for us to ask two important questions:

1) If the advance of time is probabilistic in nature it can 'stop'. It might be that the particles in the box only bump into each other in such a way that the density of the gas remains the same, so that the gas does not spread throughout the box. This is very improbable, but it is theoretically possible. Does that mean that, in such situations, time itself stops? One might argue that what we have described (entropy increase) happens *within* time, and that whenever entropy-increase stops, time keeps flowing. In such a view, our considerations about entropy say nothing about time itself.[8] In other words: *is there time when there is no change?*
2) Is the direction of time defined by entropy an *objective* direction? To be able to answer that question, we need to agree on the nature of probability.

Regrettably, there is no agreement on the nature of probability. Whether probability should be understood as an objective state of affairs or as a subjective degree of belief (perhaps representing knowledge about a physical state of affairs) is a matter of dispute.[9] Therefore, it is a moot question whether the direction of time defined by entropy increase is an *objective direction*.

## 2. Relativity

In the first quarter of the 20th century, Albert Einstein came with a radical reformation of Newton's theory. He proposed the *Theory of General Relativity*: the laws of physics are the same for all observers regardless of their state of motion with respect to each other. The consequences of this new view were breathtaking. In the words of one of Einstein's contemporaries:

> *"Henceforth, space by itself, and time by itself, are doomed to fade away into mere shadows, and only a kind of union of the two will preserve an independent reality."*[10]

### Spacetime

This union is the concept of spacetime. Unlike Newton's fixed background-time, spacetime can be curved, which happens near massive objects. Newton's gravity can be understood as the curvature of spacetime: planets rotate around the sun like rolling beads circling the center of a pot. The pot is created by the mass of the sun which bends spacetime. That is why sometimes it is written that the basic idea of relativity theory is that mass tells spacetime how to curve, while spacetime tells mass how to move. *Time in relativity is not a fixed but a flexible background*.

---

[7] It might seem as if we are attacking strawmen here. We are criticizing a definition of time that we have formulated ourselves. We can only say that the definition presented here is not our own.

[8] our considerations also give rise to the opposite question. Not only "is there time when there is no change?" but also "Could it be that there is change if there is no time?" (moving particles, but no change in entropy)

[9] *The Interpretation of Probability: Still an Open Issue?*, Maria Carla Galavotti; Philosophies 2 (3):20 (2017). Defining an objective concept of probability runs into problems with the 'principle of indifference' being a subjective principle.

[10] Address to the 80th Assembly of German Natural Scientists and Physicians, (Sep 21, 1908), reprinted in: The Principle of Relativity. Original Papers by A. Einstein and H. Minkowski (New York, 1923).

### Relativity of Simultaneity

But there is another important difference between time according to Newton and time in Einstein's theory. In Einstein's new theory, the rate at which time flows depends on the velocity of the observer – moving clocks undergo *time dilation*. Because of this, different observers see things happen at different moments. Since there is no agreement on which events are simultaneous, this result of Einstein's theory is called the *relativity of simultaneity*[11].

### Presentism no longer intuitive

We recall that presentism seemed like a very reasonable/intuitively appealing view if we take causality into account; we assume that only the present, not the past and the future, is real because we can causally interact with it (we are able to influence it). The relativity of simultaneity makes presentism less intuitively appealing because it clashes with the way we think about causes. A cause always precedes its effect. However, because of the relativity of simultaneity, different observers will disagree on which of two events was first, and therefore which might have been the cause and which the effect.[12] This means that the present for one observer will be the past or the future for another observer. *Presentism is no longer intuitively appealing* because we wouldn't know whose present is the real thing.

The empirical success (successful predictions) of Einstein's relativity theory is commonly taken to be a compelling argument in favor of *eternalism*: past, present, and future are all equally real. In this view there needs to be a psychological explanation for our sensation of passing time (in terms of the workings of the brain).[13]

So where does this leave us? The debate about the nature of time has significantly changed: relativity theory suggests that we live in a universe in which the flow of time is an illusion (the eternalist view), while it yields severe constraints on a psychological theory of time. An empirical standpoint suggests that psychological time is merely change and the relativity of simultaneity suggests that time does not have a clear direction.

### 3. Quantum Theory

Another theory that came up early in the 20th century was the quantum theory. Despite being a radically novel theory, we will see that it might revive the intuitive appeal of presentism. In a certain sense, it brings us closer to Newton's original idea about time.

To understand what quantum theory is all about, we start by looking at a game of billiards. Suppose we want to describe the regularities of a bunch of billiard balls on a billiard table (an example of such a regularity would be the conservation of momentum). We can describe these regularities in terms of the balls and their velocities, but that leaves us wondering *why* there are these velocities. We might add invisible actors (players of the billiard game) to our description, who – despite being invisible – cause the motions of the balls and therefore make it possible to predict what is going to happen.

---

[11] The relativity of simultaneously was already part of Einstein's special relativity theory published in 1905.

[12] Time dilation cannot temporally swap cause and effect. However, it *can* make the time interval between cause and effect seem very small. Clear insight about what is the effect and what is the cause requires us to say which of the two is first, so time dilation does present a difficulty for causal thinking.

[13] Is that even possible? Perhaps it is like the spokes on a wheel that seem to be standing still in a video clip when the time between different movie shots is exactly equal to the time it takes the wheel to rotate from one spoke to the next (in reverse this would create the illusion of time flow when actually there is none).

We may think of Newton's theory as a description in terms of only balls and a billiard table, while quantum theory introduces many invisible concepts that make the theory into a better theory (in terms of the accuracy of predictions). Particles and objects are no longer masses the size of a point that follow Newton's laws, but they are now represented by a wave function that follows the Schrödinger-equation.

Important for our argument is the fact that in quantum theory a particle or object no longer has a specific position at all times, but only when it is observed.[14] The wave function of a particle or object obeys Schrödinger's equation until it is observed – when an observation is made, the wavefunction collapses onto a specific point. After observation, what we are left with corresponds approximately to the description that Newton would have given (which is why, in daily life, Newtonian physics is used to describe what is going on).

Why is this important for our story? Because the collapse of the wavefunction happens everywhere at the same time. That is quite different from what relativity theory said, where it depends on the state of motion of the observer whether two events are judged to be simultaneous. Through the collapse of the wavefunction (instead of an instantaneous gravitational influence such as in Newton's theory), quantum theory opens a backdoor to the idea of a present that is the same for everyone – again we can define a global 'now'.

### Does quantum theory save presentism?
It seems, then, that quantum theory saves the intuitive idea behind presentism. But that is not the end of the story. Again, we look at things from the empirical perspective (focusing on the observable). In quantum theory, just as in Newton's theory, we see only things with a specific position – no matter how useful or accurate quantum theory is, we do not see wavefunctions or their collapse. The collapse of the wavefunction is introduced to explain why we see specific positions instead of a wavefunction when we make observations. So, the assumption of collapse serves to connect the quantum picture with actual scientific practice. However, the collapse of the wavefunction is not the only way in which the quantum picture can connect to actual scientific observations.

Another way to explain why we do not see wavefunctions is the *many worlds interpretation of quantum theory (MWI)*. In any situation where there are probabilities, according to this interpretation, the world splits up in as many different worlds – parallel realities – as there are possibilities. For example, when a die is thrown, the world splits into six different parallel realities. In each of those one of the six possible outcomes is realized.

---

[14] take a moment to let this sink in. Everything that you have ever seen in your life tells you that things have a specific position even when they are seen by nobody (they are *somewhere*). It sounds preposterous to assume that a person that you see walking in the street no longer has a specific position if you can't see her because she is passing a tree (normally you would say that she *does* have a specific position, namely behind the tree). Yet quantum theory states that particles behave very differently: if we can't observe them, they do not have a specific position.

When it was first proposed this interpretation met with ridicule, which may be understandable because it is very far from the interpretation that physicists were used to. Of course, that in itself is not an argument against the idea of many worlds.[15] If we admit – if only for the sake of the argument – that the MWI is a possible interpretation of quantum theory, we can no longer be sure that there is a collapse of the wave function, and therefore presentism once again loses its intuitive appeal.

Does this make eternalism into the preferred option? It seems worthwhile to look at our line of reasoning here: Are we willing to concede that the flow of time is a psychological phenomenon because it might be the case that parallel realities exist? Are we willing to give up age-old ideas about something we can see all around us because of the mere possibility of some wildly speculative idea about transcendental realities that we can never interact with?

## 4. Epilogue: quantum gravity/string theory

In the previous paragraphs we have seen the different arguments for both sides of the presentism/eternalism-debate. Let us end with a brief remark on how this debate continues if theoretical physics develops further.

Our current understanding of theoretical physics stands on a two-sided pedestal. On one side we have relativity theory, which is very helpful in predicting how large objects (such as planets and stars) will move. On the other side there is quantum theory which is very good at predicting what happens at a very small scale (such as inside atoms). These two theories both work very well on their respective scales but do not fit together smoothly.[16]

### The End of Time?

The foremost candidate-theories for the unification of quantum theory and relativity theory are

1) quantum gravity, and
2) string theory

Physicists around the world are trying to work out these two theories; the theories have not yet reached their final form, while there is no consensus about which of the theories is most promising. However, it is already evident that a certain equation – the Wheeler-DeWitt equation – plays an important role in both theories. And in the Wheeler-DeWitt equation there is no longer need for a time parameter.[17] It

To conclude, in the course of our paper it has become clear that physics has a lot to say about the concept of time – but it cannot say anything with certainty. It seems that the statement "time does not exist" is not entirely wrong, but we should be very careful with its interpretation. What do we mean with the word 'time'? Do we mean that Newton' time, as a rigid, non-interacting background to change, does not exist? How can we ever justify the statement that something with which nothing interacts does not exist? What do we mean when we say that something 'exists'? Is it an emergent property of how our brains work or is it an objective fact about reality? Do these two options contradict each other?

---

[15] Why should the universe be intuitive or understandable for *us*?

[16] We already talked about the 'specific position': in quantum theory particles have no specific position as long as they are not observed, yet macroscopic objects (like stars and planets) have a specific position even when nobody looks at them.

[17] According to the theoretical concept *AdS/CFT-correspondence*, it seems that time is an "emergent" phenomenon: bendy, curvy space-time and the matter within it are a hologram that arises out of a network of entangled qubits; much as the three-dimensional environment of a computer game is encoded in the classical bits on a silicon chip.

## References

1. Dainton, B., "Space and time", McGill-Queen's University Press; 2nd ed. edition (2010).
2. Einstein, A., "Zur Elektrodynamik bewegter Körper" ["On the Electrodynamics of Moving Bodies"]. *Annalen der Physik* (1905).
3. Galavotti, M.C., "The Interpretation of Probability: Still an Open Issue?" *Philosophies* (2017)
4. Newton, I., "The Mathematical Principles of Natural Philosophy", tr. Andrew Motte, London 1729. reissued as "The Principia", Amherst, NY: Prometheus (1995).
5. Rovelli, C., "The Order of Time", Penguin Books (2019)